\documentclass{article}
\usepackage[utf8]{inputenc}
\usepackage{graphicx}
\usepackage{url}
\usepackage{float}
\usepackage{authblk}

\title{COVID-19 in Mexico: A Network of Epidemics}
\author{Guillermo de Anda-Jáuregui$^{1,2}$ }
\date{\small
    $^1$ Computational Genomics Division, National Institute of Genomic Medicine, Mexico City, Mexico\\%
    $^2$Cátedras Conacyt Para Jóvenes Investigadores, National Council on Science and Technology, Mexico City, Mexico\\[0.5ex]%
}

\begin{document}

\maketitle

\section*{Abstract}

Mexico, like the rest of the world, is currently facing the The COVID-19 pandemic. Given the size of its territory, the efforts to contain the disease have involved both national and regional measures.  
For this work, the curves of daily new cases  of each municipality reported by the federal government were compared. We found that 114 municipalities form a large network of statistically dependent epidemic phenomena. Based on the network's modular structure, these 114 municipalities can be split into four distinct communities of coordinated epidemic phenomena. These clusters are not limited by geographical proximity. These findings can be helpful for public health officials for the evaluation of past strategies and the development of new directed interventions. \\

\section{Introduction}

Mexico reported its first imported case of COVID-19 in late February \cite{SSA12020}. Since then, COVID-19 has extended throughout the Mexican territory, with over 90,000 accumulated cases and over 10,000 confirmed deaths by June 2020 \cite{SSA12020b} \\

The federal government reported a major re-conversion project to increase its hospital capacity \cite{SSA12020c}. Due to the lack of pharmacological treatments against SARS-CoV-2, Mexico, like the rest of the world, resorted to the use of Non Pharmacological Interventions (NPI) to manage the spread of the disease; these efforts were branded as the "National Period of Healthy Distance" ("Jornada Nacional de Sana Distancia," JNSD), originally planned to last from March 23rd to April 30th, and then extended through May 30th \cite{Presidencia2020}. \\

An important aspect of Mexico's response to the COVID-19 pandemic is that social distancing was not enforced at the individual level. Instead, social distancing was promoted through the closing of non-essential economic activities. At the end of the JNSD, a transition strategy for the gradual reactivation of economic activities was presented, which is to be evaluated at the state level. \\

This regionalization of the COVID-19 response makes sense given the size of the mexican territory and the heterogenous distribution of its population. With this in mind, thinking of the epidemic as a single phenomenon throughout the Mexican territory may not be as adequate as thinking about many parallel epidemic processes, some of which can be interconnected through complex social dynamics. Strategies such as metapopulation models \cite{gleam} attempt to reconstruct these large-scale epidemiological networks at a global scale; however, the detailed description that these require is not readily available for the Mexican territory.  \\

In this work, a data-driven approach is used to extract this latent structure of interdependences between epidemic processes at the municipal level. An information-theoretic measure of statistical dependence is used to reconstruct a network connecting municipalities with similar behaviours in terms of their daily new cases counts. We identify that 114 municipalities are part of an interconnected network of epidemic dynamics. Furthermore, this network is composed of four modules with higher within-connectivity. Importantly, these clusters are not necessarily connected through geographical proximity, indicating that there are perhaps other sociodemographic and behavioural processes that guide these epidemic phenomena. \\

\section{Methods}

\subsection{Data acquisition}

Data used for this work is obtained from the Mexican federal government open data platform \cite{DatosMX}; Data was accessed on 2020/06/18. \\

Daily new cases were counted for each municipality (based on the patient's residence), considering the reported symptom onset day. Only confirmed cases were considered for this analysis.  Since there is a documented delay in case reporting due to the long chain of data capture \cite{Castaneda2020}, this analysis was limited to the dates between 2020/03/01 and 2020/05/31 (the end of the JNSD). For comparison purposes, daily new case counts were normalized over the total population of each municipality (based on the 2015 inter-census count) \cite{inegi2015}. \\

\subsection{Mutual information calculation}

For this work, the time series of normalized daily new cases associated to each municipality was used. To evaluate the statistical dependence between each new cases curve, mutual information (MI) was used as a similarity measure \cite{cover2012elements}. Time series were discretized and MI values were calculated using the \textit{infotheo} package for R \cite{infotheo}. \\

\subsection{Network construction and analysis}

Having the complete set of MI values for each municipality pair, a network was constructed by connecting those municipalities that have an MI value equal or higher to a threshold of 0.55. The resulting network was analyzed using the \textit{igraph} R package \cite{igraph}.  In particular, module (also known as community) detection was done using the Louvain community detection algorithm \cite{Blondel_2008}. \\

\section{Results and discussion}

\subsection{114 municipalities form a network of coexisting epidemic phenomena}

The federal government database contains records of 125577 new cases in the time period studied, distributed in 1623 municipalities. This curve is a reflection of the underlying epidemic process occurring within the community. Therefore, similarity between epidemic processes may be captured by the statistical similarity of these curves. In figure 1, the distribution of MI values is shown using a heatmap. We can observe that statistical dependence between municipalities is relatively rare.\\

\begin{center}
\begin{figure}[H]
\includegraphics[scale=0.75]{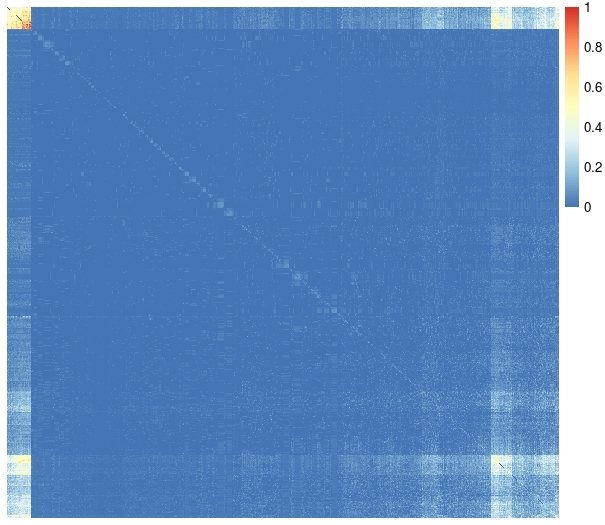} 
\caption{Heatmap of MI values between municipalities}
\end{figure}    
\end{center}



These observations are transformed into a network, using a conservative MI threshold of 0.55 to consider a link between municipalities. Analyzing the component distribution of the network, we find that at this level of significance, 114 municipalities are part of the only connected component, while the rest of the municipalities do not exhibit statistical dependence with any other. Table 1 contains basic network descriptors of the main component; a visualization of this network is found in figure 2. \\

\begin{table}[H]
\caption{Main component descriptors}
\centering
\begin{tabular}{cc}
\hline
descriptor & value \\ 
  \hline
nodes & 114 \\ 
edges & 1622 \\ 
mean\_k & 28.5 \\ 
density & 0.25 \\ 
clustering\_coefficient & 0.65 \\ 
average\_path\_length & 1.89 \\ 
   \hline
\end{tabular}
\end{table}

\begin{center}
\begin{figure}[H]
\includegraphics[scale=0.5]{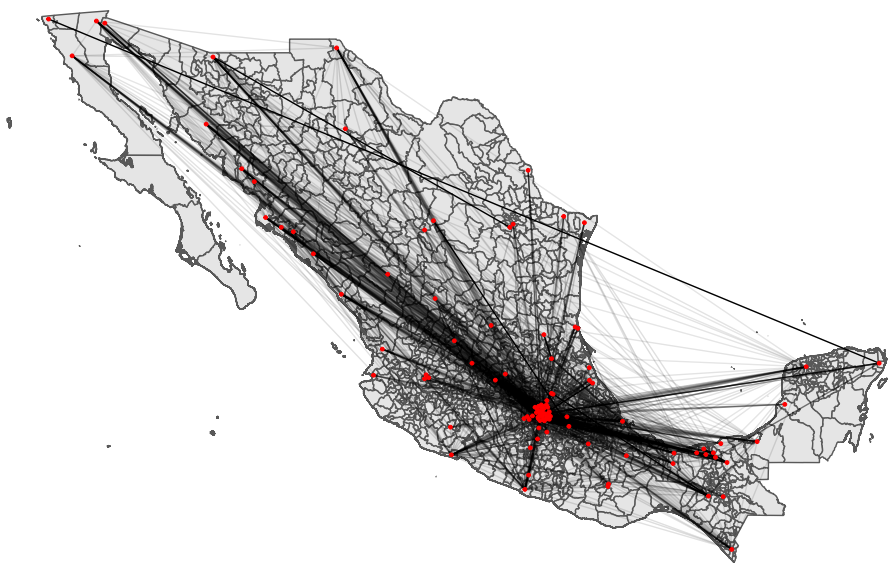} 
\caption{Network of municipalities with statistically dependent COVID-19 epidemic processes}
\end{figure}    
\end{center}

We identify that the 10 highest ranked municipalities by degree include five boroughs of Mexico City, Ecatepec and Toluca in the State of Mexico, and the ports of Veracruz, Lázaro Cárdenas, and Acapulco. Although seven of this municipalities are in the Mexican Altiplano, it is interesting to find three port cities that are not geographically nearby: two in the Pacific Coast, and one in the Gulf of Mexico. \\

\begin{table}[H]
\centering
\caption{Highest ranked municipalities by degree}
\begin{tabular}{llr}
  \hline
state & municipality & degree \\ 
  \hline
Ciudad de México & Azcapotzalco & 77 \\ 
  Ciudad de México & Gustavo A. Madero & 77 \\ 
  Ciudad de México & Coyoacán & 75 \\ 
  Veracruz de Ignacio de la Llave & Veracruz & 74 \\ 
  México & Toluca & 73 \\ 
  Ciudad de México & Venustiano Carranza & 72 \\ 
  Guerrero & Acapulco de Juárez & 72 \\ 
  México & Ecatepec de Morelos & 72 \\ 
  Michoacán de Ocampo & Lázaro Cárdenas & 70 \\ 
  Ciudad de México & Milpa Alta & 69 \\ 
   \hline
\end{tabular}
\end{table}

\subsection{The network modular structure reveals 4 distinct coordinated dynamics}

The 114 municipalities in the main component of the network contain 97926 of the total cases recorded in the studied period. Using the Louvain community detection algorithm, we identified that these network has a modular structure, shown in Figure 3. A possible interpretation of this is that each of these modules has some underlying social, demographic and environmental properties that somehow coordinate the epidemic dynamic within these municipalities. Interestingly enough, these interdependencies are not limited to contiguous zones; in fact, geographically adjacent municipalities can belong to different clusters; such is the case of the boroughs of Mexico City. In what follows, a brief description of each of these modules is provided. \\

\begin{center}
\begin{figure}[H]
\includegraphics[scale=0.5]{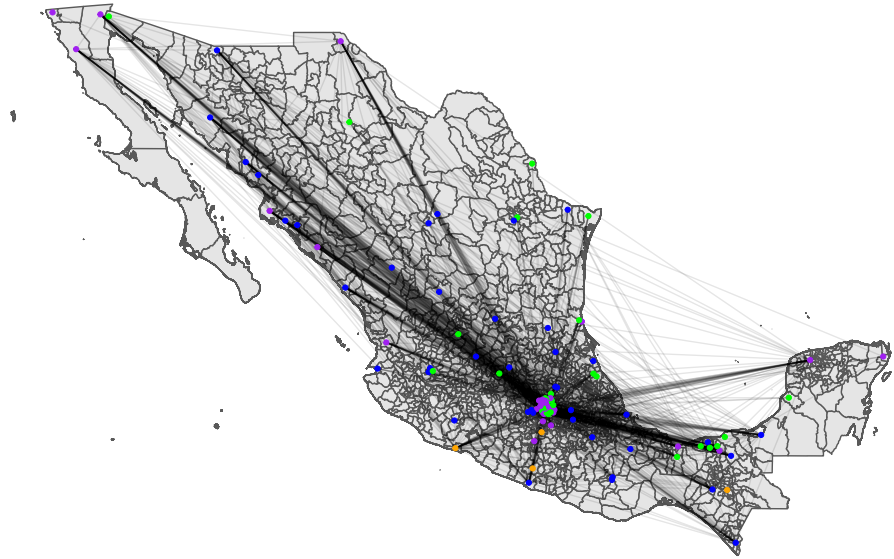} 
\caption{Modules of the COVID-19 network of epidemics}
\end{figure}    
\end{center}

\subsubsection{Module One}

\begin{center}
\begin{figure}[H]
\includegraphics[scale=0.5]{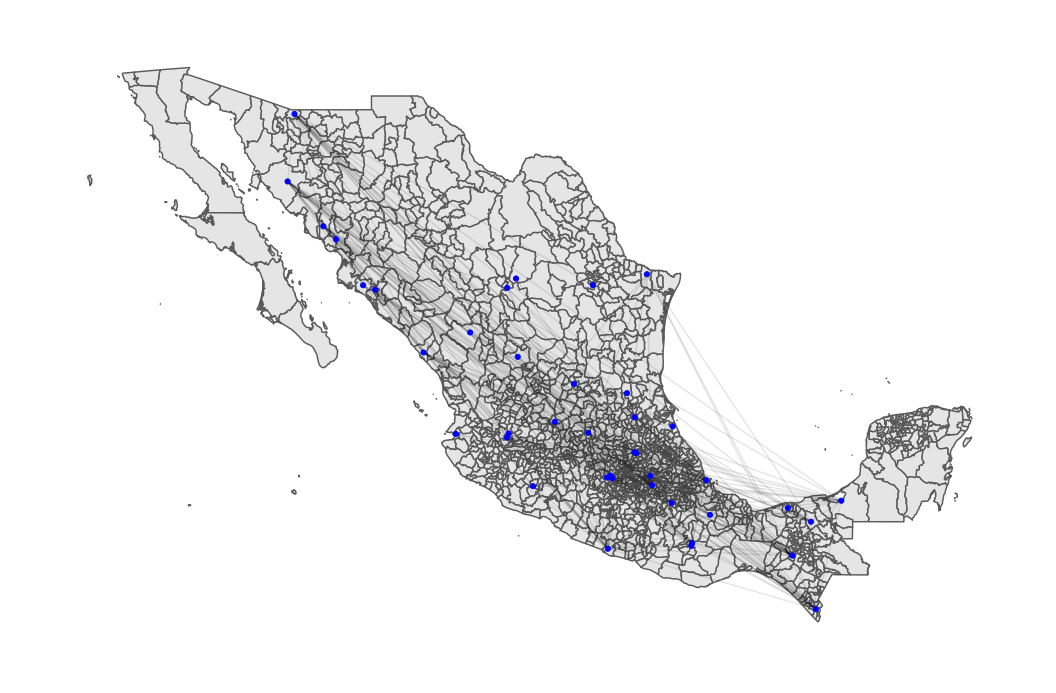} 
\caption{Subnetwork of municipalities belonging to Module One}
\end{figure}    
\end{center}

Module One is comprised of 41 municipalities distributed in 22 states (as seen in table 3). These municipalities concentrate 24304 cases of COVID-19. It is interesting to note that, although this is the largest module in terms of represented municipalities, it does not contain any of the boroughs of Mexico City. It does include, however, the capital cities of the states of Chiapas, Durango, Hidalgo, Jalisco, the State of Mexico, Nuevo Leon,  Oaxaca, Puebla, Querétaro, San Luis Potosí, Sonora, Tlaxcala, and Veracruz. In fact, in Figure 4, it can be appreciated that the municipalities that comprise this module are geographically distant. 

\begin{table}[H]
\centering
\caption{Module One}
\begin{tabular}{lr}
  \hline
state & no.municipalities \\ 
  \hline
Campeche &   1 \\ 
  Chiapas &   2 \\ 
  Coahuila de Zaragoza &   2 \\ 
  Durango &   1 \\ 
  Guanajuato &   1 \\ 
  Guerrero &   1 \\ 
  Hidalgo &   2 \\ 
  Jalisco &   3 \\ 
  México &   3 \\ 
  Michoacán de Ocampo &   1 \\ 
  Nuevo León &   1 \\ 
  Oaxaca &   3 \\ 
  Puebla &   2 \\ 
  Querétaro &   1 \\ 
  San Luis Potosí &   3 \\ 
  Sinaloa &   3 \\ 
  Sonora &   4 \\ 
  Tabasco &   2 \\ 
  Tamaulipas &   1 \\ 
  Tlaxcala &   1 \\ 
  Veracruz de Ignacio de la Llave &   2 \\ 
  Zacatecas &   1 \\ 
   \hline
\end{tabular}
\end{table}

\subsubsection{Module Two}

\begin{center}
\begin{figure}[H]
\includegraphics[scale=0.5]{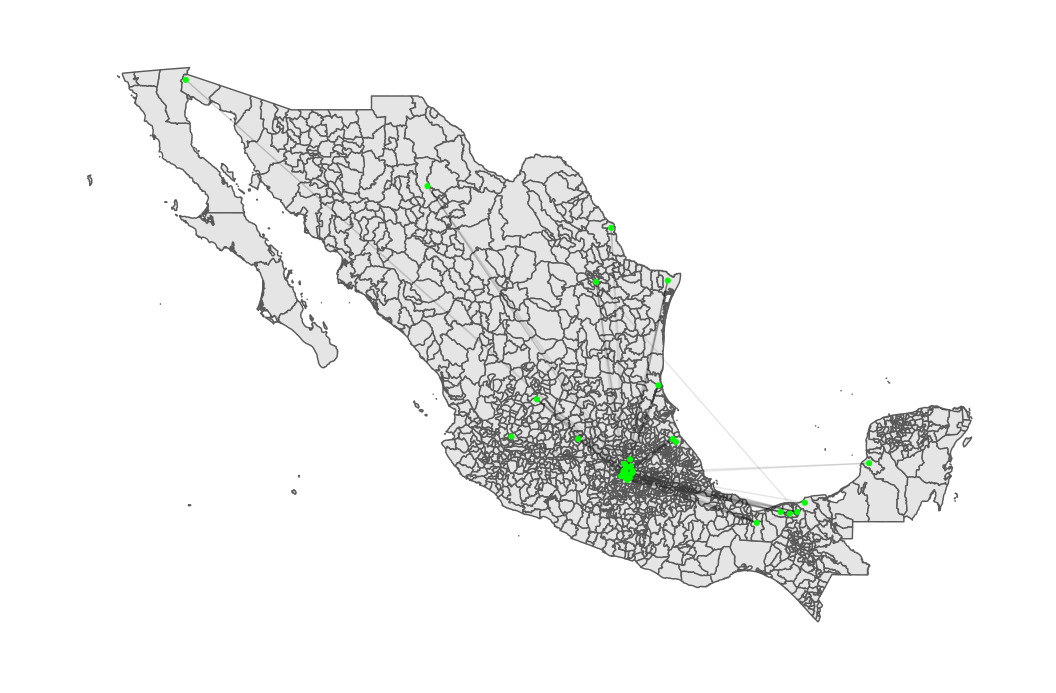} 
\caption{Subnetwork of municipalities belonging to Module Two}
\end{figure}    
\end{center}

Module Two is comprised of 35 municipalities distributed in 13 states (as seen in table 4). These municipalities concentrate 28269 cases of COVID-19. This module contains the capital cities of Campeche, and Chihuahua. It also contains nine boroughs of Mexico City, and eight municipalities of the state of Mexico. 

\begin{table}[H]
\centering
\caption{Module 2}
\begin{tabular}{lr}
  \hline
estado & no.municipalities \\ 
  \hline
Aguascalientes &   1 \\ 
  Campeche &   1 \\ 
  Chihuahua &   1 \\ 
  Ciudad de México &   9 \\ 
  Guanajuato &   1 \\ 
  Hidalgo &   1 \\ 
  Jalisco &   1 \\ 
  México &   8 \\ 
  Nuevo León &   1 \\ 
  Sonora &   1 \\ 
  Tabasco &   4 \\ 
  Tamaulipas &   3 \\ 
  Veracruz de Ignacio de la Llave &   3 \\ 
   \hline
\end{tabular}
\end{table}

\subsubsection{Module Three}

\begin{center}
\begin{figure}[H]
\includegraphics[scale=0.5]{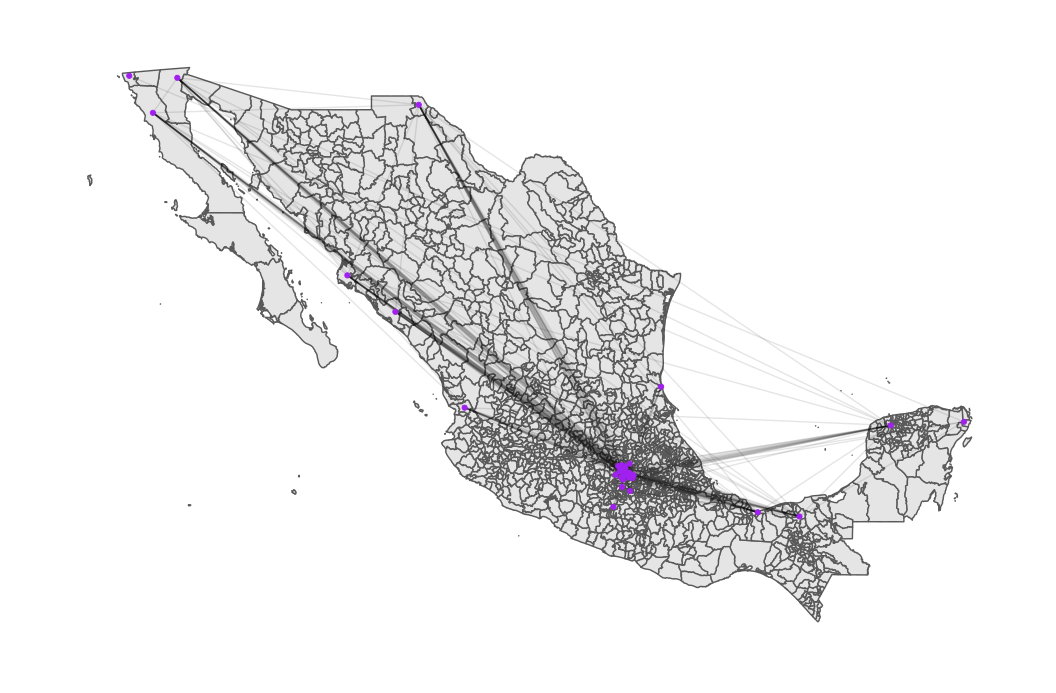} 
\caption{Subnetwork of municipalities belonging to Module Three}
\end{figure}    
\end{center}

Module Three is comprised of 34 municipalities distributed in 13 states (as seen in table 5). These municipalities concentrate 43421 cases of COVID-19, making it the module that concentrates the largest number of cases. This module contains 6 boroughs of Mexico City, 13 municipalities in the State of Mexico, the municipalities of Cuautla and Cuernavaca in Morelos, the border cities of Tijuana, Mexicali, and Juarez, the capital cities of the western states of Nayarit and Sinaloa, and the the largest cities in the Yucatan Peninsula: Mérida and Cancún. 

\begin{table}[H]
\centering
\caption{Module 3}
\begin{tabular}{lr}
  \hline
estado & n \\ 
  \hline
Baja California &   3 \\ 
  Chihuahua &   1 \\ 
  Ciudad de México &   6 \\ 
  Guerrero &   1 \\ 
  México &  13 \\ 
  Morelos &   2 \\ 
  Nayarit &   1 \\ 
  Quintana Roo &   1 \\ 
  Sinaloa &   2 \\ 
  Tabasco &   1 \\ 
  Tamaulipas &   1 \\ 
  Veracruz de Ignacio de la Llave &   1 \\ 
  Yucatán &   1 \\ 
   \hline
\end{tabular}
\end{table}

\subsubsection{Module Four}

\begin{center}
\begin{figure}[H]
\includegraphics[scale=0.5]{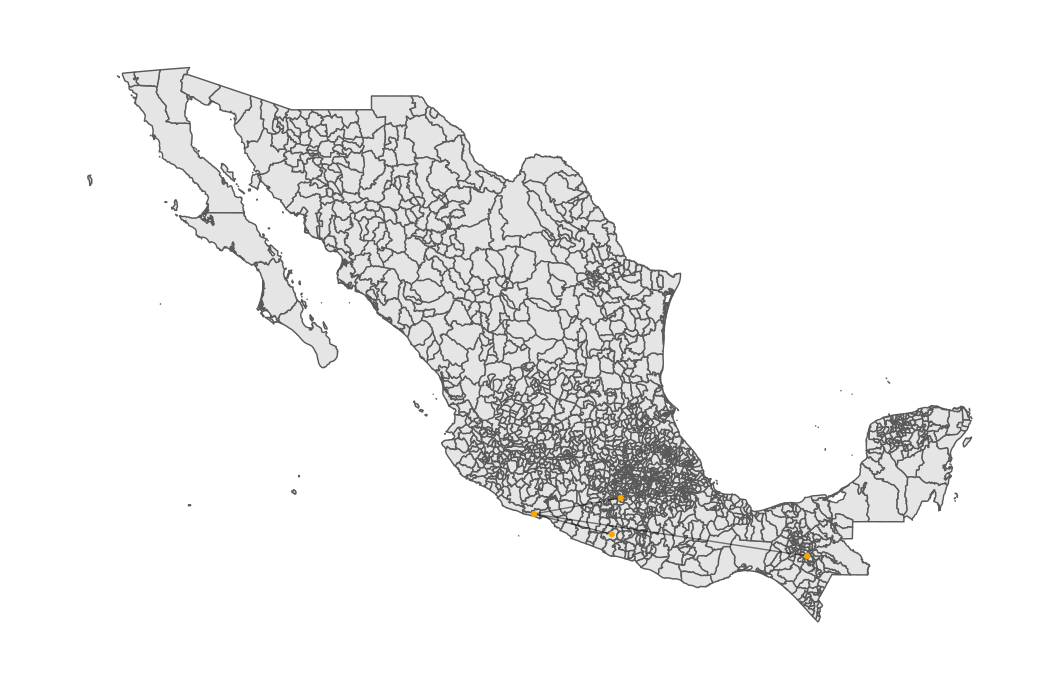} 
\caption{Subnetwork of municipalities belonging to Module Four}
\end{figure}    
\end{center}

Module Four is the smallest of the network's partitions. It is composed of only four municipalities in the states of Chiapas, Guerrero,  Michoacán and Morelos, spread across the Mexican south. Interesingly, these four municipalities are bound together through the port of Lázaro Cárdenas, Michoacán; previously mentioned as one of the most connected nodes in the network. 

\begin{table}[H]
\centering
\caption{Module 4}
\begin{tabular}{lr}
  \hline
estado & n \\ 
  \hline
Chiapas &   1 \\ 
  Guerrero &   1 \\ 
  Michoacán de Ocampo &   1 \\ 
  Morelos &   1 \\ 
   \hline
\end{tabular}
\end{table}






\section{Conclusions}

Through this network exploratory analysis of the COVID-19 cases reported by the Mexican government, we are able to identify patterns of epidemic phenomena behaving in a somewhat coordinated fashion. It is shown that these correlations in the daily cases are not mediated solely by geographical effects, with connections spanning the length of the Mexican republic. We can hypothesize that these connections can reflect similarities in the heterogeneous transmission dynamics occurring in each locality \cite{althouse2020stochasticity}.
As such, an argument can be made that the underlying mechanisms that drive this coordination can be a combination of societal, demographic, economic, and other phenomena with subtle, yet intricate connections.\\

It is important to mention the widely discussed limitations in terms of the COVID-19 Mexican open datasets. The reporting delay has been already addressed and was accounted for in the initial selection of the time period for the analysis. Another known limitation is that confirmatory testing is heavily biased towards symptomatic cases that match an "operative definition" \cite{defoper}.\\ 

However, since the data aggregation is mediated by the federal government, there is some methodological consistency that allows for somewhat fair comparisons between geographical locations; something that is not possible in countries where such centralized mechanisms are not in place; or for general comparison between nation states in in which confounding factors abound, making it unfeasible \cite{oreilly2020}. However, it is expected that logistical limitations may skew detection of cases toward larger urban areas; such bias correction is currently outside the scope of this work. \\

This manuscript provides evidence that the complex system that is the COVID-19 epidemic in Mexico can be described as several coexisting epidemics, which may exhibit similarity to other concurrent epidemic phenomena regardless of geographical distance. As an exploratory analysis, providing mechanistic explanations of these observations is currently beyond the scope of this work. However, an appeal is made to both policy makers and the general public of consider a much more granular approach to both mitigation and reopening strategies, as well as to personal risk assessment. 

\bibliographystyle{unsrt}
\bibliography{biblio}

\section*{Acknowledgements}

The author would like to thank Irving Morales for providing a file with processed inter-census population data; Jesús Espinal-Enríquez for providing computational resources; Enrique Hernández-Lemus for discussion of this manuscript. 

\section*{Code availability}

Code used for this work is available at:\\ 
https://github.com/guillermodeandajauregui/mxCovid19\_nw\\

\end{document}